\documentclass[letterpaper,12pt]{article}
\usepackage[version=3]{mhchem} 
\usepackage[utf8]{inputenc}
\usepackage[english]{babel}
\usepackage{helvet}
\usepackage{graphicx}
\usepackage{color}
\usepackage{float}
\usepackage{amssymb}
\usepackage{booktabs}
\usepackage{enumerate}
\usepackage{upgreek}
\usepackage[normalem]{ulem}
\usepackage{fancyhdr}
\usepackage{titlesec}
\usepackage[
    backend=biber,
    style=phys,
    sorting=none,
    giveninits=true,
    articletitle=true
]{biblatex}

\AtBeginBibliography{\emergencystretch=2em}
\usepackage{xcolor}
\newcommand{\rev}[1]{#1}

\begin{document}

\thispagestyle{empty}
\begin{center}
    \large
    \textbf{Structural disorder and critical voltage scaling in $\ce{Al}$/$\ce{AlOx}$/$\ce{Al}$ Josephson junction arrays}
    
    \vspace{0.4cm}
    \normalsize
    Amanuel M. Berhane$^{1,2,*}$, Susan Coppersmith$^{2}$, Emma Mitchell$^{1}$, Timothy Duty$^{2}$
    
    \vspace{0.4cm}
    $^{1}$\textit{CSIRO Manufacturing, PO Box 218, Lindfield, 2070, NSW, Australia}\\
    $^{2}$\textit{School of Physics, University of New South Wales, 2052, NSW, Australia}
    
    \vspace{0.4cm}
    $^{*}$\textit{amanuel.berhane@csiro.au}
    
    \vspace{0.4cm}
       
\end{center}
The insulating state of one-dimensional Josephson junction (JJ) arrays is governed by collective charge dynamics and disorder-induced pinning, resulting in a finite critical voltage under dc bias. \rev{Here, we investigate the influence of fabrication-induced structural defects on the critical-voltage scaling of small-capacitance Aluminium--Aluminium oxide--Aluminium ($\mathrm{Al}/\mathrm{AlO_x}/\mathrm{Al}$) JJ arrays. Controlled variation of the aluminium evaporation rate produces pronounced changes in grain morphology and room-temperature junction resistance. Despite these substantial structural modifications, the normalized critical-voltage scaling is preserved, demonstrating that the collective transport behaviour is remarkably robust against this class of fabrication-induced defects. In contrast, the deliberate introduction of nanoscale gaps into the junctions introduces additional junction-to-junction structural variations that systematically modify the normalized scaling behaviour. Likewise, \textit{in situ} post-fabrication oxidation alters the scaling coefficient while preserving the functional form of the scaling law, indicating that the collective transport is sensitive to specific classes of structural modifications. These results establish which fabrication-induced structural defects influence the collective transport in insulating $\mathrm{Al}/\mathrm{AlO_x}/\mathrm{Al}$ Josephson junction arrays, providing new insight into the role of fabrication-induced structural disorder and practical guidance for the design of future Quantum Phase Slip (QPS) devices.}

\pagebreak

\section{Introduction}

One-dimensional (1D) arrays of Josephson junctions provide a well-controlled platform for studying quantum charge transport in the presence of strong interactions and disorder. In the insulating regime, where the Josephson energy ($E_{\mathrm{J}}$) is small compared to the Cooper-pair charging energy ($E_{\mathrm{CP}}$), transport is governed by Coulomb blockade and collective charge dynamics, leading to a finite critical voltage ($V_{\mathrm{C}}$) under dc bias \cite{likharev1985theory,haviland1991observation}. Such systems are of interest both for exploring quantum phase-slip (QPS) physics \cite{bouchoule2025platforms,haviland2001quantum,ergul2013phase,PhysRevB.92.045435,pop2010measurement} and for potential applications in metrology \cite{crescini2023evidence,guichard2010phase} and superconducting circuit design \cite{randeria2024dephasing,koliofoti2023compact,osborne2024symplectic}. 

In practice, $\ce{Al}$/$\ce{AlOx}$/$\ce{Al}$ junction arrays are generally understood to realise a Bose-glass ground state, where disorder pins collective charge modes setting the magnitude of $V_{\mathrm{C}}$ \cite{giamarchi2017theory}. Previous experimental and theoretical studies have shown that in small-capacitance one-dimensional arrays, $V_{\mathrm{C}}$ exhibits a characteristic scaling behaviour with the ratio $E_{\mathrm{J}}/E_{\mathrm{CP}}$ over a wide range of plasma frequencies \cite{haviland2001quantum,chow1998length}. This scaling has been observed across various experimental regimes and interpreted within collective depinning and pinned Luttinger-liquid descriptions of disordered Josephson junction arrays \cite{PhysRevLett.119.167701}. In these theories, the dominant source of disorder is random island offset charges, which determine the collective charge dynamics underlying the Bose-glass state.

\rev{Bard \textit{et al.} further showed that fluctuation in the local quantum phase-slip (QPS) amplitude,
$y\propto e^{-\alpha K_\mathrm{0}}$, constitute an additional microscopic disorder mechanism within the renormalisation-group description of the superconductor--insulator transition in 1D Josephson junction arrays \cite{bard2017superconductor,mukhopadhyay2023superconductivity}. Here, $\alpha$ is a screening-length-dependent numerical coefficient and $K_{0}$ is the bare Luttinger parameter (phase stiffness), given by 
\begin{equation}
K_{0}
=
\frac{\pi}{\Lambda}\sqrt{\frac{E_{\mathrm J}}{2E_\mathrm{CP}}},
\end{equation}
where $\Lambda=\sqrt{E_{0}/E_{\mathrm{CP}}}$
is the electrostatic screening length and $E_{0}$ is the charging energy associated with capacitive coupling to the ground. Because the QPS amplitude depends exponentially on the bare Luttinger parameter, even modest spatial variations in the local junction energies, together with random stray charges, can produce significant fluctuations in the QPS amplitude. Consequently, both random island offset charges and fluctuations in the QPS amplitude influence the renormalisation-group flow that determines whether the system evolves towards the superconducting or insulating phase \cite{feldman2025quantum,kuzmin2019quantum}. Because the critical voltage provides an experimental measure of the collective insulating state, changes in the microscopic disorder mechanisms are expected to be reflected in the critical-voltage scaling behaviour.}

Fabrication-related processes such as evaporation geometry, oxidation conditions, and film morphology can introduce atomic and nanoscale defects at the junction barrier and island--junction interfaces. \rev{ Such structural defects are expected to introduce junction-to-junction variations in the local parameters and modify the microscopic disorder mechanisms governing collective charge transport, including both random offset charges and fluctuations in the local quantum phase-slip (QPS) amplitude. It therefore remains unclear which classes of fabrication-induced structural defects simply renormalise the critical-voltage scaling coefficients and which fundamentally modify the scaling behaviour itself.}

In this study, we systematically investigate the effect of \rev{fabrication-induced structural defects} on the critical voltage scaling behaviour of 1D $\ce{Al}$/$\ce{AlOx}$/$\ce{Al}$ Josephson junction arrays. By fabricating families of arrays using controlled variations of the double-angle evaporation process, including evaporation geometry, $\textit{in situ}$ post-fabrication oxidation, and evaporation rate, we \rev{introduce distinct classes of structural defects} while keeping junction sizes nominally comparable. Combining room-temperature conductivity measurements, scanning electron microscopy, and low-temperature (20 mK) transport measurements, we show that while evaporation rate and post-fabrication oxidation strongly affect junction resistance, the functional form of the critical voltage scaling remains robust. \rev{Post-fabrication oxidation primarily modifies the scaling prefactors while preserving the functional form of the scaling behaviour.} On the other hand, the inclusion of nanoscale gaps at the island--junction interfaces introduces \rev{additional structural defects} that significantly alter the observed scaling behaviour.

\section{Experimental methods}

 The one-dimensional arrays studied here consist of 1000 $\ce{Al}$/$\ce{AlOx}$/$\ce{Al}$ junctions (\(N = 1000\))
 connected in series. The arrays were fabricated using the standard electron-beam lithography with a bilayer resist (MMA/ARP) lift-off process and thermal double-angle shadow evaporation. Most devices were fabricated on 10 $\times$ 10 mm$^{2}$ $\ce{SiO2}$ (300 nm)/$\ce{Si}$ substrates, while a limited number were fabricated on intrinsic $\ce{Si}$ and $\ce{Si3N4}$(300 nm)/$\ce{Si}$ substrates of the same size. Metal deposition was carried out at base pressure below 2$\times$10$^{-7}$ mbar, with a nominal $\ce{Al}$ thickness of 30 nm for both evaporation steps. The $\textit{in situ}$ static oxidation between the two evaporations was used to control the junction plasma frequency ($\omega_{\mathrm {p}}$) and was controlled by oxidation pressure ($\textit{P}_{\mathrm {ox}}$) and time ($\textit{t}_{\mathrm {ox}}$).
 
\begin{figure}[ht]
\centering\includegraphics[scale=21]{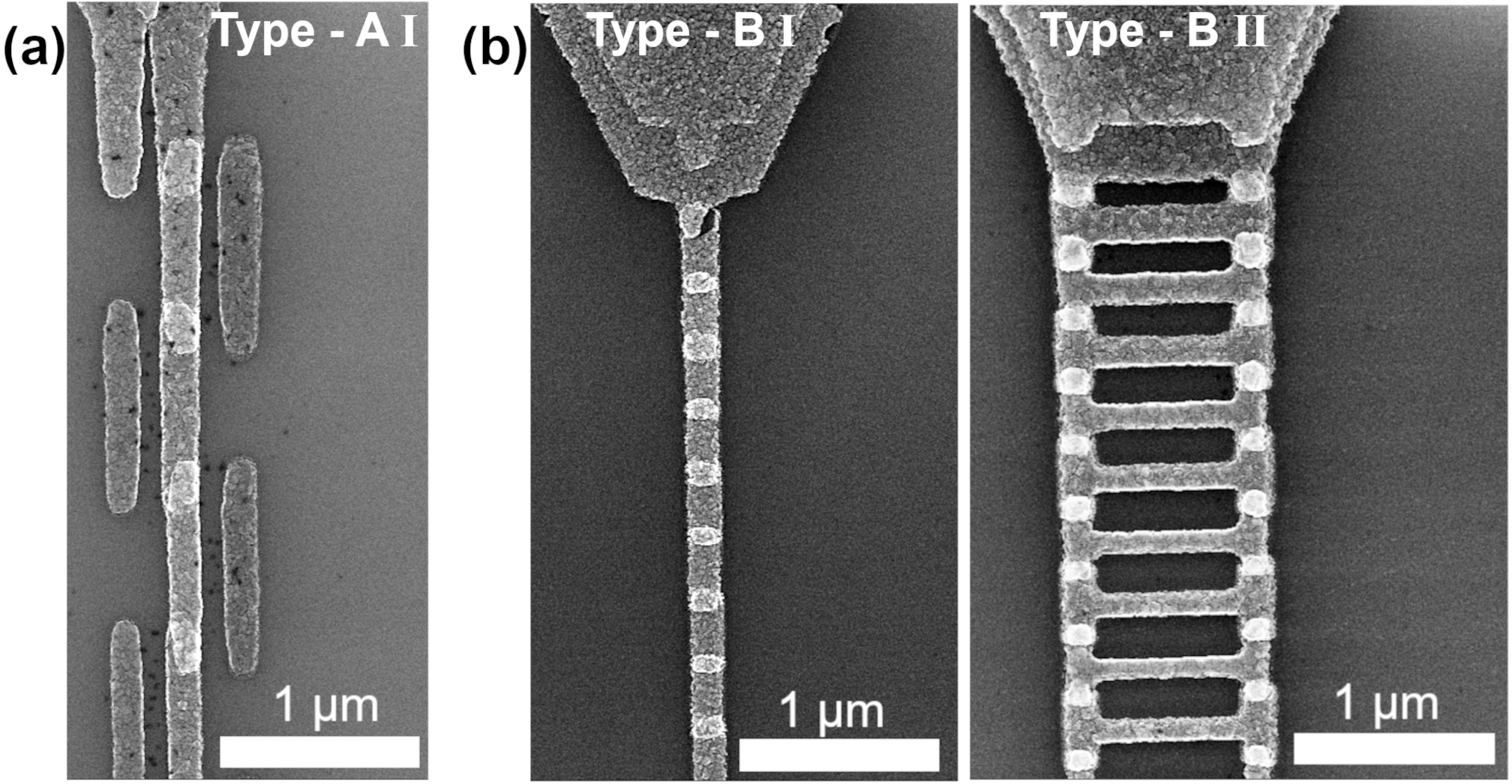}
\caption{Representative SEM images showing examples of Type-A and Type-B Josephson junction arrays.  
(a) Type-AI array consisting of 800~nm-long islands with uniform island geometry and the presence of excess metal strips. The nominal junction area of this family of arrays is \(100 \times 100\)~nm\(^2\).  
(b) Type-B arrays fabricated without excess metal strips. The left panel shows a single-junction chain (Type-BI) consisting of 400~nm-long islands with a uniform island width of 200~nm and a nominal junction area of \(100 \times 200\)~nm\(^2\). The right panel shows a SQUID array (Type-BII) fabricated with the same island and junction dimensions as the Type-BI chains. The Type-BI and Type-BII arrays were fabricated on the same chip under identical processing conditions.
}
\label{fig:Figure1}
\end{figure}

Following Ref.~\cite{PhysRevLett.119.167701}, junction and island sizes were changed by assigning different electron-beam exposure doses to the lithography mask, allowing arrays with systematically varying junction areas to be fabricated on the same chip. In addition, by employing mask designs that allow evaporation both parallel and perpendicular to the chain direction, two families of arrays (Type-A and Type-B) were fabricated. Figure~\ref{fig:Figure1} shows representative scanning electron micrographs of Type-A [Fig.~1(a)] and Type-B [Fig.~1(b)] arrays. The Type-B family comprises two designs: Type-BI, a single-junction chain analogous to the Type-A design but without excess metal strips, and Type-BII, a SQUID array fabricated on the same substrate during the same deposition run as the Type-BI arrays. 

Transport measurements were performed on the fabricated arrays at both room temperature and low temperature. Room-temperature conductivity was measured using a Keithley Cascade probe station (4200SCS/C) in a two-point configuration. Since the applied voltage drops predominantly across the tunnel barriers, the two-point configuration provides a reliable measure of junction transport. Low-temperature dc current--voltage characteristics (IVCs) were obtained using Keithley K6430 and Yokogawa GS200 source-measure units (SMUs) by cooling the devices to 20~mK in a BlueFors LD400 cryogen-free dilution refrigerator. \rev{The samples were mounted on the mixing-chamber stage with appropriate thermal anchoring and filtered measurement wiring to ensure efficient thermalisation during the low-temperature measurements} (see Ref.~\cite{PhysRevLett.119.167701,cedergren2015parity} for experimental details).  

Low-temperature current-voltage characteristics were measured over both the large-bias (\(e V \gg 2 N \Delta\)) and small bias (\(e V \ll 2 N \Delta\)), where \(V\) and \(\Delta\) denote the applied bias voltage and the superconducting gap, respectively. The large-scale IVC data were used to determine the normal junction resistance at 20~mK (\(R_{\mathrm{J}}\)), from which the Josephson energy was extracted using \(E_{\mathrm {J}} = \frac{\Delta}{2} \frac{R_\mathrm{Q}}{R_{\mathrm{J}}}\), with \(\Delta = 210~\mu\mathrm{eV}\) and \(R_\mathrm{Q} = 6.45~\mathrm{k}\Omega\) the superconducting quantum resistance. The Cooper-pair charging energy (\(E_{\mathrm{CP}}\)) was determined by fitting the normal conductance at large bias voltages, where \(E_{\mathrm{CP}} = \frac{4 e V_{\mathrm{off}}}{N}\) and \(V_{\mathrm{off}}\) is the voltage intercept. The small-scale IVCs were collected using subgap excitation, from which the critical voltage \(V_{\mathrm {C}}\) was extracted. \(V_{\mathrm {C}}\) is defined as the threshold voltage at which the measured current exceeds the noise floor by at least three orders of magnitude (see Refs.~\cite{PhysRevLett.119.167701,cedergren2015parity} for details). 

\section{Results and Discussion}
\subsection{\rev{Fabrication-induced defects} and critical-voltage scaling}

Across all devices studied, variations in evaporation rate, oxidation procedure, and evaporation geometry produced pronounced changes in room-temperature transport. Despite the substantial differences in the tunnelling characteristics of the junctions, the functional scaling behaviour of $V_{\mathrm{C}}$ with the ratio $E_{\mathrm{J}}$/$E_{\mathrm{CP}}$ varied from remaining essentially unaffected to exhibiting markedly different characteristics. \rev{The observed changes in room-temperature junction resistance may arise from systematic variations in the average junction parameters, such as the average tunnel-barrier thickness, differences in the extent of fabrication-induced structural defects, or a combination of both. The influence of these structural modifications on the low-temperature critical-voltage scaling is examined in the following sections.}
\subsubsection{Evaporation rate, grain size, and morphology}
\label{sec:Weak disorder}

The evaporation rate of the aluminium films strongly affects their grain size and hence surface roughness \cite{nik2016correlation,fritz2019optimization,oh2025correlating}, leading to pronounced variations in $R_{\mathrm{j}}$. Slow evaporation produces coarse-grained, rough films that are associated with greater oxidation and higher junction resistance, whereas faster evaporation yields smoother films with reduced oxidation and lower $R_{\mathrm{j}}$. The influence of the evaporation rate on the grain size and morphology of the films is evident in the SEM images and is accompanied by systematic changes in the RT conductivity.

In this study, the morphology of Type-AI $\ce{Al}$/$\ce{AlOx}$/$\ce{Al}$ junction chains was varied by fabricating arrays at different aluminium evaporation rates, as monitored using a Quartz Crystal Microbalance (QCM). Representative SEM images of junctions fabricated at different evaporation rates are shown in Fig.~2(a--c). At the slowest evaporation rate (0.3~\AA/s), the films exhibit a coarse-grained polycrystalline morphology, with large grooves between coalescing grains and a relatively rough surface, as shown in Fig.~2(a). In contrast, increasing the evaporation rate to 4~\AA/s produces a fine-grained polycrystalline film with a much smoother morphology, as shown in Fig.~2(c).

Each \(10 \times 10\) mm\(^2\) substrate contained four sets of ten arrays prepared in a single deposition run: two sets with a nominal junction areas of \(100 \times 100\) nm\(^2\) and two sets with a nominal junction area of \(100 \times 200\) nm\(^2\). The replicate sets were positioned at the opposite corners of the substrate. After fabrication, the substrate was diced into four \(3 \times 3\) mm\(^2\) dies, each containing ten arrays with systematically increasing junction area. Figure~2(d) shows the average room-temperature single-junction resistance on a semi-logarithmic scale as a function of electron-beam exposure dose (400--1300~\(\mu\)C/cm\(^2\)) for two sets of Type-AI \(\ce{Al}/\ce{AlOx}/\ce{Al}\) chains (\(N = 1000\)) fabricated at evaporation rates of 0.3, 1, and 4~\AA/s. Increasing the exposure dose systematically increases the nominal junction area, with the nominal \(100 \times 100\) nm\(^2\) junction obtained at the lowest exposure dose of 400~\(\mu\)C/cm\(^2\). The average \(R_j\) as a function of exposure dose for the additional pair of ten arrays (\(N = 1000\)) with nominal junction areas of \(100 \times 200\) nm\(^2\) is shown in Fig.~2(e) for the three evaporation rates. For arrays fabricated at 0.3~\AA/s, the average \(R_j\) is approximately fifteen times larger than that of arrays prepared at 1 and 4~\AA/s at the same exposure dose. The modest difference in the average \(R_j\) between the 1 and 4~\AA/s devices is comparable to the typical run-to-run variability of the oxidation process. The characteristic non-linear dependence of \(R_j\) on exposure dose arises from the non-linear relationship between the resist opening area and the applied electron-beam exposure dose.
\begin{figure}[H]
\centering\includegraphics[scale=0.8]{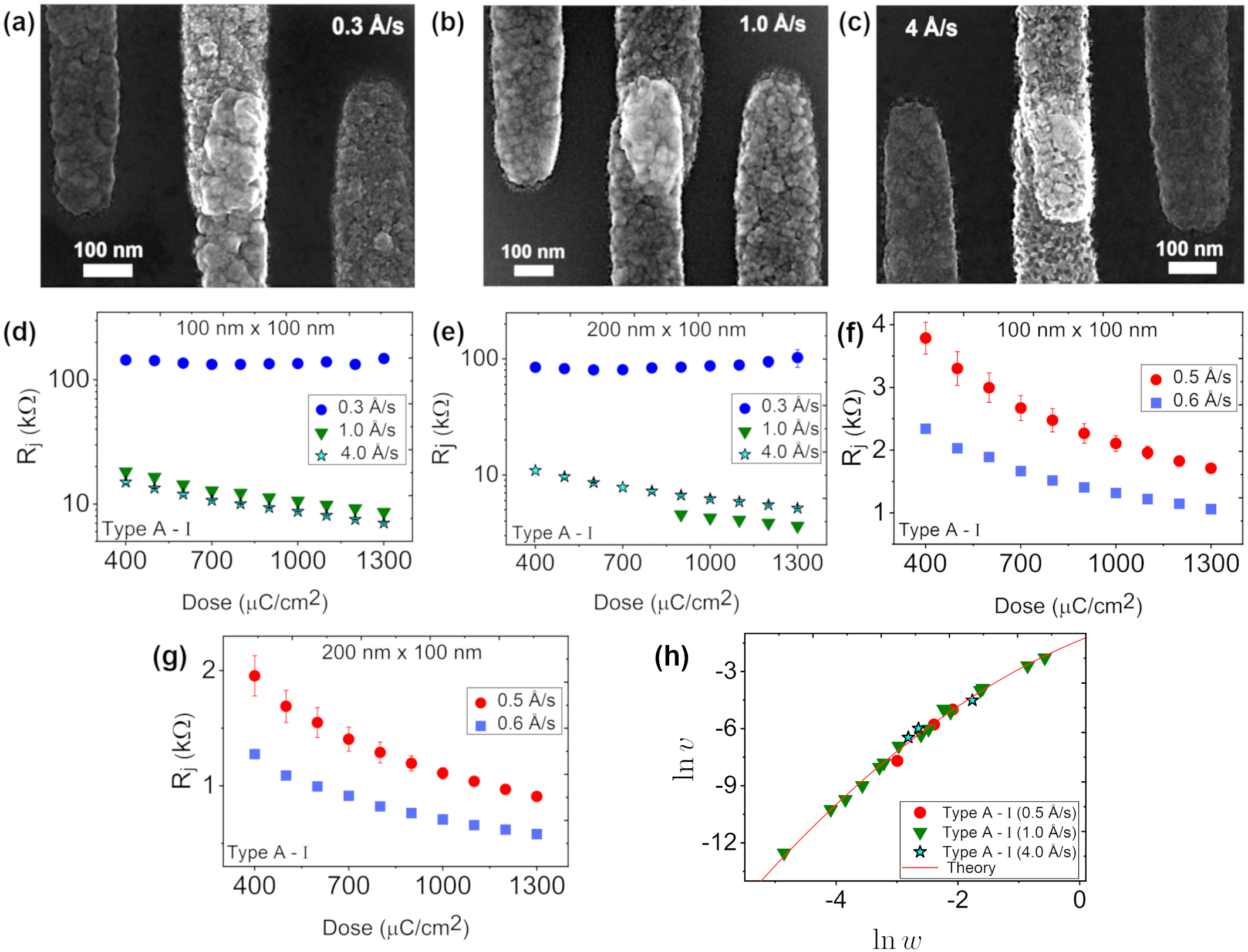}
\caption{SEM images of Type-AI junctions fabricated at evaporation rates of 0.3~\AA/s (a), 1.0~\AA/s (b), and 4.0~\AA/s (c), illustrating the systematic evolution of film morphology with deposition rate. Average room-temperature single-junction resistance, \(R_j\), as a function of electron-beam exposure dose for Type-AI arrays (\(N = 1000\)) with nominal junction areas of \(100 \times 100\)~nm\(^2\) (d) and \(100 \times 200\)~nm\(^2\) (e), fabricated at evaporation rates of 0.3, 1.0, and 4.0~\AA/s \rev{under identical oxidation conditions (\(P_{\mathrm{ox}} = 1~\mathrm{mbar}\), \(t_{\mathrm{ox}} = 30~\mathrm{s}\))}. The missing data points in (e) for exposure doses between 400 and 800~\(\mu\)C/cm\(^2\) correspond to failed devices. \rev{Panels (f) and (g) compare the dose dependence of \(R_j\) for arrays fabricated at evaporation rates of 0.5 and 0.6~\AA/s under identical oxidation conditions (\(P_{\mathrm{ox}} = 2 \times 10^{-2}~\mathrm{mbar}\), \(t_{\mathrm{ox}} = 30~\mathrm{s}\)), highlighting the strong sensitivity of junction resistance to small changes in evaporation rate in the low-rate regime. (h) Normalized critical-voltage scaling of the Type-AI arrays shown in (d)--(g), demonstrating that despite the large variation in room-temperature junction resistance induced by evaporation rate, the functional scaling remains unchanged. The solid red line is a fit based on the pinned Luttinger-liquid theory \cite{PhysRevLett.119.167701}.}
}
\label{fig:Figure2}
\end{figure}
The higher junction resistance at slow evaporation rates is consistent with enhanced oxidation of the rougher, coarse-grained films, which possess a larger effective surface area. This interpretation is supported by ellipsometry measurements performed on separately deposited 30~nm Al films oxidised under identical conditions. While approximately 44\% oxide fraction (effective composition) is extracted for the film deposited at 0.5~\AA/s, only 29\% is estimated for the film deposited at 1~\AA/s. For evaporation rates above \(\sim 1\)~\AA/s, the dependence of \(R_j\) on grain size saturates, as evidenced by the comparable conductivities of arrays fabricated at 1 and 4~\AA/s [Figs.~2(d) and 2(e)]. This saturation suggests that, above an evaporation rate of approximately 1~\AA/s, further changes in film morphology produce comparatively little change in the effective oxidation behaviour of the junctions \cite{cabrera1949theory,gorobez2021growth}. \rev{In contrast, for evaporation rates below around 1~\AA/s, \(R_j\) depends strongly on grain size. As shown in Figs.~2(f) and 2(g), slightly reducing the evaporation rate from 0.6 to 0.5~\AA/s while maintaining  the same growth conditions increases the room-temperature junction resistance by approximately 50$\%$ for both \(100 \times 100\) nm\(^2\) and \(100 \times 200\) nm\(^2\) wide junctions. Furthermore, after accounting for differences in oxygen exposure (defined as a the product of $\textit{P}_{\mathrm o \mathrm x}$ and $\textit{t}_{\mathrm o \mathrm x}$), arrays fabricated at 0.5~\AA/s exhibit room-temperature junction resistances approximately five times larger than those of arrays fabricated at 1~\AA/s, highlighting the strong sensitivity of room temperature transport to film morphology and grain size.}

Arrays fabricated at the slowest evaporation rate of 0.3~\AA/s exhibit an anomalous room-temperature conductivity behaviour in which the resistance increases with increasing junction area. To investigate this effect, an additional set of arrays was prepared at the same evaporation rate but with a substantially reduced oxidation time. Although the overall single-junction resistance \(R_j\) decreases, the anomalous area dependence persists. Moreover, the reduction in \(R_j\) does not scale linearly with oxidation time: \(R_j\) decreases by only a factor of two despite a reduction of more than an order of magnitude in oxidation time under similar partial pressures. These observations suggest that, at the lowest evaporation rate, the characteristic grain dimensions become comparable to the junction dimensions, causing the effective tunnel-barrier properties to depend nontrivially on the local grain configuration. The resulting structural inhomogeneity may account for the observed departure from the conventional decrease of \(R_j\) with increasing junction area.

The critical voltages of several Type-AI arrays fabricated at evaporation rates of 0.5, 1, and 4~\AA/s and spanning different values of \(g=E_{\mathrm J}/E_{\mathrm{CP}}\) were extracted from dc current--voltage characteristics measured at 20~mK. To determine whether variations in evaporation rate modify the depinning behaviour, \(V_{\mathrm C}\) was analysed as a function of the Bloch bandwidth

\begin{equation}
W = 16 \sqrt{\frac{E_{\mathrm J}E_{\mathrm{CP}}}{4\pi}}
(2g)^{1/4}
e^{-\sqrt{32g}}.
\end{equation}

Figure~2(h) presents the resulting dimensionless scaling relation, where

\begin{equation}
\upsilon=\frac{eV_{\mathrm C}}{N\hbar\omega_{\mathrm p}},
\qquad
w=\frac{W}{\hbar\omega_{\mathrm p}}.
\end{equation}
Despite the pronounced dependence of the room-temperature junction resistance on aluminium evaporation rate, the normalized critical voltages collapse onto a common scaling relation over the investigated range of plasma frequencies. Following the pinned-Luttinger-liquid theory developed in Ref.~\cite{PhysRevLett.119.167701}, the scaled critical voltage and Bloch bandwidth are related by

\begin{equation}
\upsilon = a w^{\alpha},
\end{equation}

where \(a\) is a fitting parameter and the exponent is given by \(\alpha (w)= 4/(3 - 2K_{0}(w))\).

\rev{A fit to the dense 1~\AA/s dataset, shown by the solid red curve in Fig.~2(h), indicates that the logarithmic slope, $\alpha(w)$, evolves continuously with increasing Bloch bandwidth. To quantify this evolution directly from the measurements, five-point moving linear regressions were performed on the 1~\AA/s dataset. Each regression yields a local logarithmic slope,
\begin{equation}
\alpha_{\mathrm{eff}}
=
\frac{d\ln\upsilon}{d\ln w},
\end{equation}
which was subsequently interpreted within the pinned-Luttinger-liquid framework to obtain the model-inferred effective Luttinger parameter,
\begin{equation}
K_{\mathrm{eff}}
=
\frac{1}{2}
\left(
3-\frac{4}{\alpha_{\mathrm{eff}}}
\right).
\end{equation}
Although the 0.5 and 4~\AA/s datasets contain too few measurements to independently determine the evolution of \(K_{\mathrm{eff}}\) over the full disorder range, their measurements overlap with the dense 1~\AA/s dataset over the interval -2.82 $\lesssim$ $\ln w$ $\lesssim $-2.09.}

\rev{To examine whether the local depinning behaviour depends on evaporation rate, all measurements within this common interval were analysed together. A joint linear regression over the interval -2.82 $\lesssim$ $\ln w$ $\lesssim $-2.09 yields $\alpha_{\mathrm{eff}}$ = 2.15 $\pm$ 0.31, corresponding to
$K_{\mathrm{eff}}$ = 0.57 $\pm$ 0.13. Within experimental uncertainty, this single effective stiffness describes the measurements obtained at evaporation rates of 0.5, 1, and 4~\AA/s over their common bandwidth. Thus, although the complete evolution of \(K_{\mathrm{eff}}\) can presently only be resolved for the dense 1~\AA/s dataset, the available data demonstrate that the local depinning response is insensitive to evaporation rate when compared at the same scaled Bloch bandwidth.}

\rev{To determine whether devices fabricated at different evaporation rates exhibit comparable microscopic phase stiffness, analogous to the common effective phase stiffness inferred from the local scaling analysis above, $K_{0}$ was independently calculated for each device using the experimentally determined electrostatic screening length. Following the gate-periodicity analysis of Cedergren \textit{et al.}~\cite{PhysRevLett.119.167701}, the screening length of the arrays was determined to be $\Lambda=5.5$. Using this value together with Eq.~(1), $K_{0}$ was calculated for each device. A comparison of devices fabricated at the three evaporation rates over comparable regions of the normalized scaling curve is summarized in Table~\ref{tab:K0comparison}. Over the common $\ln(w)$ interval, the calculated microscopic phase stiffness differs by less than $6.2\%$ between evaporation rates. Together with the common local scaling exponent, which yields a common model-inferred effective phase stiffness $K_{\mathrm{eff}}$, these results demonstrate that, despite substantial variations in room-temperature resistance and film morphology, devices fabricated at different evaporation rates exhibit essentially identical normalized critical-voltage scaling. The quantitative difference between the independently calculated microscopic phase stiffness $K_{0}$ and the model-inferred effective stiffness $K_{\mathrm{eff}}$ is consistent with renormalisation of the depinning behaviour by disorder.}

\begin{table}[t]
\rev{
\centering
\caption{Comparison of the calculated bare phase stiffness $K_{0}$ for devices fabricated at different evaporation rates. Each 0.5~\AA/s and 4~\AA/s device is compared with the nearest 1~\AA/s device in normalized Bloch bandwidth $\ln(w)$. The percentage difference is defined relative to the corresponding 1~\AA/s device.}
\label{tab:K0comparison}
\begin{tabular}{cccccc}
\toprule
Rate (\AA/s) &
$\ln(w)$&
 $K_{0}$&
Nearest $\ln(w)$
(1~\AA/s) &
$K_{0}$ (1~\AA/s) &
$\Delta K_{0}$ (\%) \\
\midrule
0.5  & -2.0861 & 0.2520 & -2.1217 & 0.2553 & -1.3 \\
    & -2.3936 & 0.2804 & -2.4822 & 0.2884 & -2.8 \\
    & -2.9971 & 0.3340 & -2.9721 & 0.3318 & +0.7 \\
\midrule
4.0  & -1.7636 & 0.2213 & -1.6316 & 0.2085 & +6.2 \\
     & -2.6484 & 0.3034 & -2.6075 & 0.2997 & +1.2 \\
    & -2.8154 & 0.3181 & -2.9721 & 0.3318 & -4.1 \\
\bottomrule
\end{tabular}
}
\end{table}

\rev{The collapse of the normalized critical-voltage scaling indicates that varying the evaporation rate predominantly modifies the spatially averaged junction properties while producing no resolvable change in the effective disorder governing collective pinning. The observed scaling collapse constrains the net influence of offset-charge disorder and fluctuations in the quantum phase-slip amplitude on the measured collective depinning response. These results therefore demonstrate that the normalized depinning characteristics are remarkably robust against substantial fabrication-induced variations in the film morphology.}

\subsubsection{\textit{In situ} post-oxidation}
\label{sec:Disorder suppression}

Having established that the fabrication-induced structural modifications associated with varying the evaporation rate do not measurably alter the critical-voltage scaling behaviour, we next investigate whether a fabrication process designed to reduce structural inhomogeneity influences the scaling behaviour. As described above, the \(\ce{AlOx}\) tunnel barrier is normally formed by \textit{in situ} oxidation of the first \(\ce{Al}\) layer at low pressure to the desired thickness, followed by deposition of the second \(\ce{Al}\) layer. The sample is then removed from the evaporation chamber for the lift-off process and subsequent room-temperature transport measurements. The results discussed in Sec.~\ref{sec:Weak disorder} correspond to arrays fabricated using this standard procedure.

A separate set of Type-AI devices was fabricated by introducing an additional oxidation step, referred to here as \textit{in situ} post-oxidation. In this process, following deposition of the second \(\ce{Al}\) layer, the samples were retained in the evaporation chamber and subjected to a second low-pressure oxidation treatment prior to lift-off. This procedure has previously been shown to suppress water-vapour absorption and improve the stability of the tunnel barrier upon exposure to ambient conditions \cite{koppinen2006effects}.

Figure~3(a) shows the average room-temperature single-junction resistance, \(R_j\), as a function of exposure dose for two \(3 \times 3\) mm\(^2\) chips, each containing ten Type-AI arrays (\(N = 1000\)) with nominal junction areas of \(100 \times 100\) nm\(^2\). One set underwent low-pressure (0.03~mbar) post-oxidation (Type-AI--\(\ce{O2}\)), while the other was fabricated using the standard process. Both sets were fabricated at an evaporation rate of 1~\AA/s. Figure~3(b) shows the corresponding measurements for arrays with nominal junction areas of \(100 \times 200\) nm\(^2\). In both cases, post-oxidised devices exhibit room-temperature junction resistances that are at least a factor of two lower than those fabricated using the standard process for comparable junction sizes.
\begin{figure}[h]
\centering\includegraphics[scale=0.78]{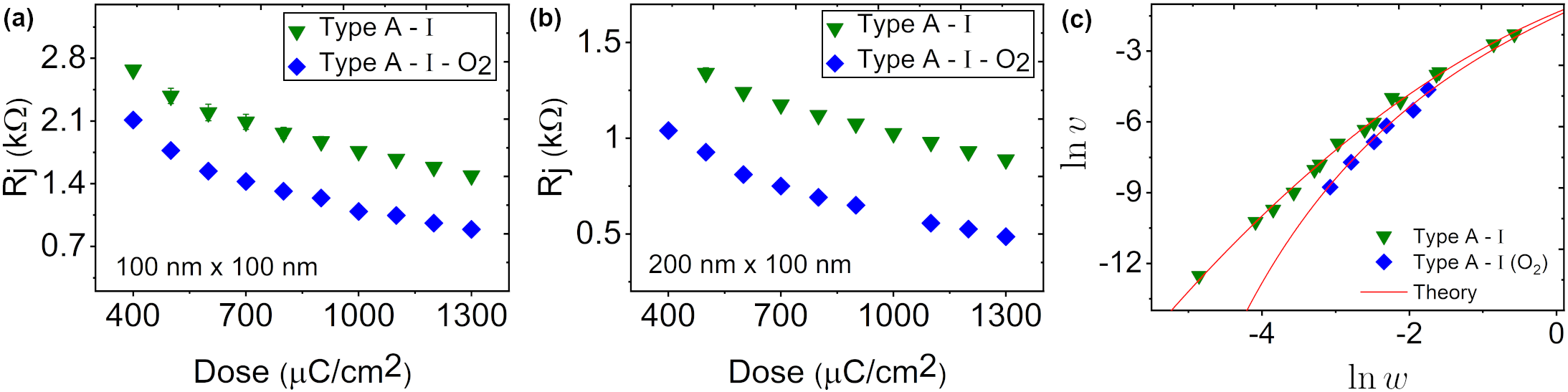}
\caption{Average room-temperature single-junction resistance \(R_j\) as a function of electron-beam exposure dose (junction area) for Type-AI arrays (\(N = 1000\)) fabricated with and without \textit{in situ} post-oxidation, with all other processing parameters kept constant. (a) Arrays with nominal junction areas of \(100 \times 100\)~nm\(^2\). (b) Arrays with nominal junction areas of \(100 \times 200\)~nm\(^2\). The missing data point at an exposure dose of 1000~\(\mu\)C/cm\(^2\) in (b) corresponds to a failed device. \rev{(c) Critical-voltage scaling plot for Type-AI arrays fabricated with and without post-oxidation. The solid red lines show fits based on the pinned Luttinger-liquid theory \cite{PhysRevLett.119.167701}.}
}
\label{fig:Figure3}
\end{figure}

This reduction in \(R_j\) is consistent with an annealing-like effect and may arise from the passivation of electrically active atomic-scale defects within the tunnel barrier. Scanning electron microscopy of post-oxidised and non-post-oxidised devices does not reveal any discernible differences in junction morphology, indicating that the observed changes are unlikely to arise from geometric variations and instead are associated with modifications to the tunnel barrier, such as changes in its defect population or effective thickness.

\rev{The low-temperature behaviour of post-oxidised Type-AI arrays is compared with that of arrays presented in Sec.~\ref{sec:Weak disorder} fabricated at the same evaporation rate. Figure~3(c) shows that the critical-voltage scaling behaviour of these two sets of arrays differs, particularly in the strong-coupling regime. Again, the moving-window regressions were performed for the
both sets of data. Over the range sampled by
the post-oxidised devices of -2.52 $\lesssim$ $\ln w$ $\lesssim$ -2.25, the local scaling exponent is
$\alpha^\mathrm{O_{2}}_{\mathrm{eff}}\simeq2.82$--$2.92$, corresponding to
$K^\mathrm{O_{2}}_{\mathrm{eff}}\simeq0.79$--$0.82$. In contrast, the non-post-oxidised
devices exhibit $\alpha_{\mathrm{eff}}\simeq2.36$--$2.41$ and
$K_{\mathrm{eff}}\simeq0.65$--$0.67$ over comparable values of $\ln w$.
Thus, in-situ post oxidation produces a clearly distinguishable reduction
in the local critical-voltage scaling and increase in the effective stiffness
inferred within the pinned-Luttinger-liquid framework. 
}

\begin{table}[t]
\centering
\rev{
\caption{Comparison of the calculated microscopic phase stiffness $K_0$ for post-oxidised and non-post-oxidised Josephson-junction chains. Each post-oxidised device is matched to the non-post-oxidised device with the nearest value of $\ln(w)$. The percentage difference is calculated relative to the matched non-post-oxidised value.}
\label{tab:K0oxidationcomparison}
\small
\begin{tabular}{ccccc}
\toprule
$\ln(w)$ &
\shortstack{$K_{0}$(Type-AI-O$_{2}$)} &
\shortstack{Nearest $\ln(w)$} &
\shortstack{$K_{0}$(Type-AI)} &
\shortstack{$\Delta K_{0}$(\%)} \\
\midrule
-3.0758 & 0.3461 & -3.2166 & 0.3528 & -1.9 \\
-2.7911 & 0.3276 & -2.9721 & 0.3318 & -1.3 \\
-2.4796 & 0.2882 & -2.4822 & 0.2884 & -0.1 \\
-2.3097 & 0.2879 & -2.2361 & 0.2659 & +8.2 \\
-1.9473 & 0.2533 & -2.1217 & 0.2553 & -0.8 \\
-1.7430 & 0.2236 & -1.6316 & 0.2085 & +7.2 \\
\bottomrule
\end{tabular}
}
\end{table}

\rev{As in Sec.~3.1.1, the bare phase stiffness $K_0$ was calculated for the post-oxidised devices (Type-AI-O$_2$) and compared with that of the Type-AI chains without post oxidation, as summarised in Table~\ref{tab:K0oxidationcomparison}. At small values of $\ln(w)$, the calculated bare phase stiffness remains comparable for the two fabrication procedures, despite the clear divergence in the critical-voltage scaling over the same range, consistent with a change in the renormalised effective stiffness. Within the theoretical framework of Bard \emph{et al.}, these observations could arise either from a reduction in the spatial fluctuations of the junction energies, which determine variation in the quantum phase-slip amplitude, or from changes in the offset-charge disorder. The latter appears less likely, since the \textit{in situ} post-oxidation treatment is not expected to increase the random offset-charge disorder. Consequently, the observed reduction in $V_{\mathrm C}$ is more naturally interpreted as resulting from reduced junction-to-junction variations in the tunnel-barrier properties, leading to a weaker effective pinning potential while preserving the functional dependence of the critical-voltage scaling.}

\subsubsection{Nanoscale gaps from evaporation geometry}
\label{sec:Enhanced disorder}
The preceding sections demonstrate that modifications to the tunnel barrier preserve the overall scaling behaviour. We now consider a qualitatively different behaviour from Type-B arrays fabricated using parallel double-angle evaporation. In this geometry, shadowing of the second \(\ce{Al}\) deposition by the first systematically produces nanoscale gaps at alternating junctions along the array. Figures~4(a) and 4(b) show magnified SEM images of the Type-BI and Type-BII arrays previously presented in Fig.~1(b). The nanoscale gaps, indicated by the red arrows, are clearly visible at every second junction along the chain.

This shadowing effect was originally identified as a limitation of the Type-B double-angle evaporation geometry by Dolan \cite{dolan1977offset}, and was later exploited to fabricate junctions with controlled nanoscale gaps \cite{kanda2005simple}. In the present study, we investigate the influence of these nanoscale gaps on the critical-voltage scaling behaviour of insulating Type-B one-dimensional arrays in comparison with the Type-A devices. As discussed below, the junction-to-junction variations in the size and morphology of the nanoscale gaps provide a plausible source of random spatial variations in the local junction energies.
\begin{figure}[h]
\centering\includegraphics[scale=21]{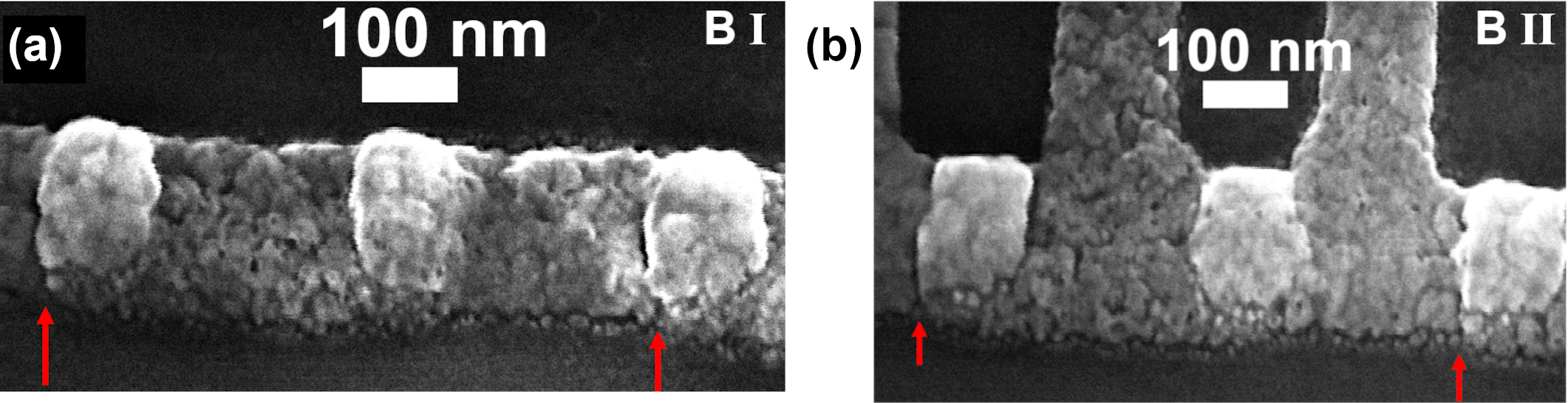}
\caption{(a) and (b) show magnified SEM images of Type-BI and Type-BII arrays, respectively, in which three representative junctions are shown, illustrating the systematic formation of nanoscale gaps at every other junction (red arrows) due to shadow evaporation in parallel direction.}
\label{fig:Figure4}
\end{figure}
\rev{It is important to distinguish the nanoscale gaps investigated here from the well-known `A/B' junction asymmetry inherent to double-angle evaporation, which is present in both the Type-A and Type-B arrays. The `A/B' effect results in a deterministic alternation of the junction and island sizes due to narrowing of lithography mask during metal deposition. By contrast, the nanoscale gaps observed in the Type-B arrays arise from shadowing during the parallel evaporation process and, to our knowledge, have received comparatively little attention in studies of Josephson-junction arrays. Moreover, the size, shape and morphology of these gaps vary appreciably from junction to junction, providing a plausible source of random spatial variations in the local junction energies.}

Differences in the transport properties of the Type-A and Type-B arrays were further examined using room-temperature conductivity measurements. Figure~5(a) shows the average single-junction resistance, \(R_j\), as a function of exposure dose (400--1300~\(\mu\)C/cm\(^2\)) for two sets of \(3 \times 3\) mm\(^2\) chips, each containing ten one-dimensional arrays (\(N = 1000\)) of Type-AI, Type-BI, and Type-BII devices with nominal junction areas of \(100 \times 100\) nm\(^2\). The room-temperature measurements show that the \(R_j\) values of the Type-BI arrays are approximately a factor of two larger than those of the Type-AI arrays, consistent with the nanoscale gaps modifying the effective tunnel-barrier properties.

Although the Type-A and Type-B arrays were fabricated in separate process runs, identical exposure doses, deposition rates, and oxidation conditions were used. The observed increase in \(R_j\) is therefore most naturally attributed to the presence of the nanoscale gaps in the Type-B geometry, which modify the effective tunnel-barrier properties. This interpretation is supported by measurements on the Type-BII (SQUID) arrays, which yield average \(R_j\) values comparable to those of the Type-AI arrays, even though the SQUID arrays comprise parallel pairs of 1000 junctions connected in series. In the absence of additional resistance associated with the nanoscale gaps, the equivalent single-junction resistance extracted from the SQUID arrays would be expected to be approximately half that of the Type-AI arrays. This behaviour is indeed observed when comparing the average \(R_j\) values of the Type-BI and Type-BII devices, which were fabricated during the same process run on a single \(10 \times 10\) mm\(^2\) \(\ce{SiO2}/\ce{Si}\) substrate before being diced into separate dies. Furthermore, the dependence of \(R_j\) on exposure dose for the Type-BI arrays exhibits substantially greater scatter than that of the Type-A arrays, consistent with junction-to-junction variations in the dimensions and morphology of the nanoscale gaps. \rev{The enhanced variability in the room-temperature resistance of the Type-B arrays is consistent with the observed junction-to-junction variations in the nanoscale-gap dimensions.}
\begin{figure}[H]
\centering\includegraphics[scale=0.74]{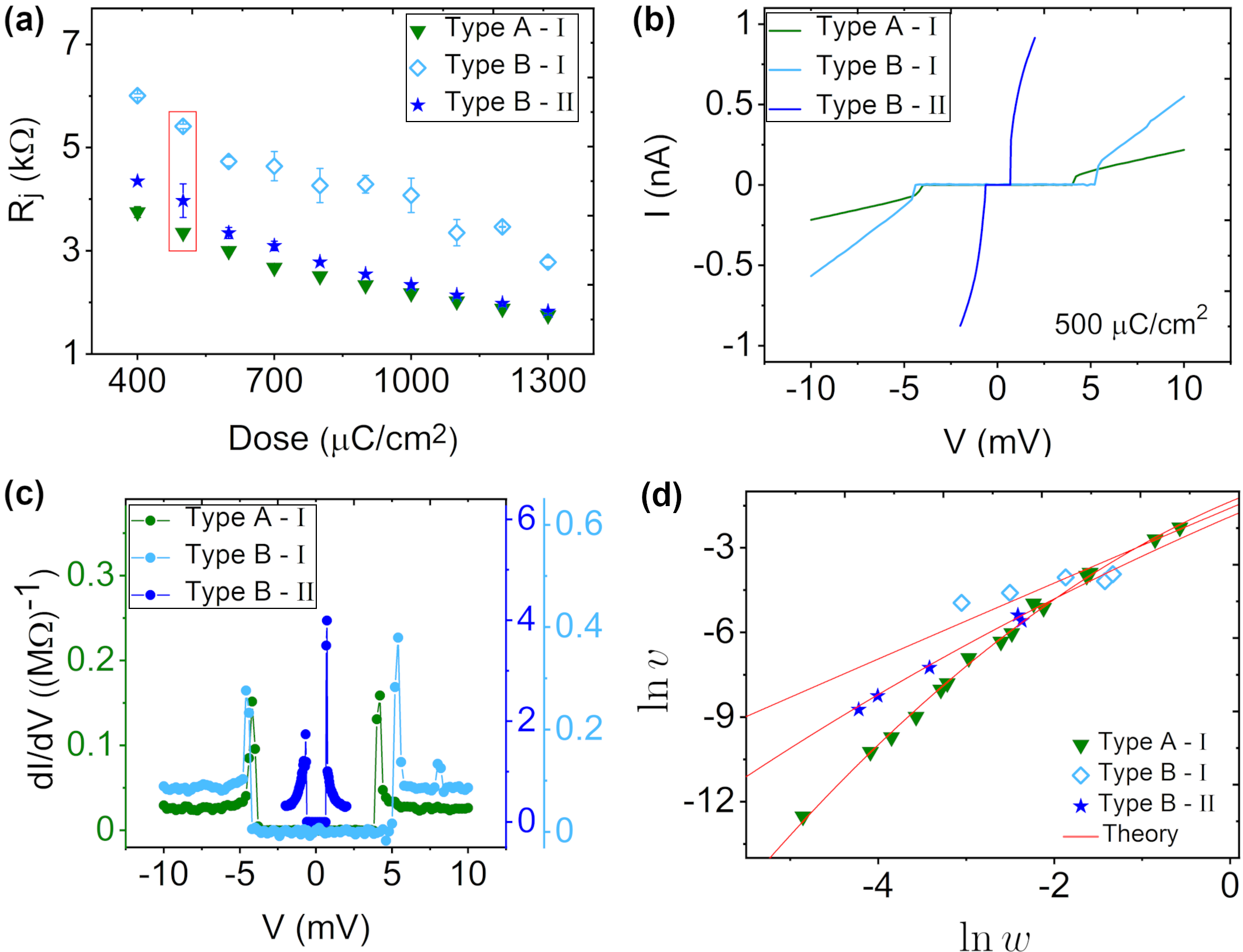}
\caption{Room-temperature single-junction resistance \(R_j\) as a function of electron-beam exposure dose (junction area) for Type-AI and Type-BI/Type-BII arrays. (b) Low-temperature (20~mK) dc current--voltage characteristics (IVCs) of three representative arrays of each type selected from the red box in (a), all fabricated at the same exposure dose of 500~\(\mu\)C/cm\(^2\) (corresponding to nominally identical junction areas).  
(c) Differential conductance traces corresponding to the IVCs in (b), highlighting the contrasting non-linear transport behaviour of the three device types.  
\rev{(d) Scaled critical voltage \(\upsilon\) as a function of the scaled Bloch bandwidth \(w\) for Type-A and Type-B devices revealing distinct scaling behaviours. The solid red lines show fits based on the pinned Luttinger-liquid theory \cite{PhysRevLett.119.167701}.}
} 
\label{fig:Figure5}
\end{figure} 
Figure~5(b) shows representative small-scale IVCs of the three arrays prepared at an exposure dose of 500~\(\mu\)C/cm\(^2\) highlighted in Fig.~5(a) (red box). In each case, the zero-current state (within the measurement noise level) persists until the applied bias reaches the critical voltage \(V_{\mathrm C}\), beyond which the system switches abruptly into a finite-current state. The corresponding differential conductance (\(dI/dV\) versus \(V\)) traces, shown in Fig.~5(c), highlight this non-linear switching behaviour. The forward critical voltages (\(+V_{\mathrm C}\)) of the Type-AI, Type-BI, and Type-BII arrays are 4.2~mV, 5.4~mV, and 0.72~mV, respectively, while the corresponding reverse critical voltages (\(-V_{\mathrm C}\)) are \(-4.2\)~mV, \(-4.6\)~mV, and \(-0.68\)~mV. The symmetric IVCs observed in the Type-A arrays and the pronounced asymmetry (diode-like behaviour) exhibited by the Type-B arrays indicate that the nanoscale gaps substantially modify the local transport properties of the junctions. The finite-current states under subgap excitation also differ markedly between the three device types. From the flat regions of the differential conductance traces, we extract conductivities of 0.03, 0.08, and 0.3~(M\(\Omega\))\(^{-1}\) for the Type-AI, Type-BI, and Type-BII arrays, respectively.

\rev{Determining the local logarithmic slope using the five-point moving linear regression [Eq.~(5)] for the Type-AI, Type-BI, and Type-BII arrays yields distinct scaling exponents. Comparing the Type-AI and Type-BI arrays over the Bloch-bandwidth interval $-4.22 \lesssim \ln(w) \lesssim -2.37$, the corresponding local logarithmic slopes are found to be $\alpha_{\mathrm{eff}}^{\mathrm{AI}}=2.36$ and $\alpha_{\mathrm{eff}}^{\mathrm{BI}}=0.54$, respectively. The markedly reduced scaling exponent for the Type-BI arrays is consistent with the introduction of nanoscale gaps, which are expected to enhance spatial fluctuations in the local junction energies and the associated quantum phase-slip amplitude, thereby strengthening the collective pinning potential. This interpretation is consistent with the substantially larger critical voltages observed for the Type-BI arrays.}

\rev{Applying Eq.~(6) to the Type-BI arrays yields the model-inferred effective Luttinger parameter $K_{\mathrm{eff}}^{\mathrm{BI}}$ = -2.18,
which is unphysical. Furthermore, fitting the Type-BI scaling data using the Cedergren model [Fig.~5(c)] results in a negative goodness-of-fit ($R^{2}<0$), with the fitted screening length reaching the imposed upper bound of the optimisation routine. The emergence of an unphysical $K_\mathrm{eff}$, together with the failure of the fit, suggests that the junction-to-junction variations introduced by the nanoscale gaps place the Type-BI arrays outside the regime over which the pinned-Luttinger-liquid description provides a physically meaningful interpretation.}

\rev{In contrast to the Type-BI arrays, the Type-BII devices yield a physically meaningful model-inferred effective Luttinger parameter of $K_{\mathrm{eff}}^{\mathrm{BII}}=0.33$, obtained by interpreting the local logarithmic slope, $\alpha_{\mathrm{eff}}^{\mathrm{BII}}=1.75$, within the pinned Luttinger-liquid framework. Over the common Bloch-bandwidth interval
$-3.05 \lesssim \ln(w) \lesssim -1.33$,
the local logarithmic slope decreases from
$\alpha_{\mathrm{eff}}^{\mathrm{AI}}=2.38$
for the Type-AI arrays to
$\alpha_{\mathrm{eff}}^{\mathrm{BII}}=1.75$
for the Type-BII arrays. The corresponding reduction in the model-inferred effective Luttinger parameter indicates that the nanoscale gaps renormalise the collective depinning response while remaining within the regime over which the pinned Luttinger-liquid description provides a physically meaningful interpretation. The distinct scaling behaviour observed between the single-junction (Type-BI) and SQUID-based (Type-BII) arrays, despite being fabricated under identical processing conditions, remains unclear. One possible explanation is that the larger Josephson energy of the Type-BII arrays reduces the sensitivity of the quantum phase-slip amplitude to local junction-to-junction parameter variations. Consequently, the effective pinning disorder experienced by the collective charge excitations would be weaker than in the Type-BI arrays, accounting for the smaller critical voltages observed in the Type-BII devices.}

\begin{table}[t]
\rev{
\centering
\caption{Comparison of the calculated bare phase stiffness $K_{0}$ for the Type-BI and Type-BII arrays. Each device is compared with the nearest Type-AI device in normalized Bloch bandwidth $\ln(w)$. The percentage difference is defined relative to the corresponding Type-AI device.}
\label{tab:K0comparison_nanogaps}
\begin{tabular}{cccccc}
\toprule
Type &
$\ln(w)$ &
$K_{0}$ &
Nearest $\ln(w)$
(Type-AI) &
$K_{0}$ (Type-AI) &
$\Delta K_{0}$ (\%) \\
\midrule
Type-BI
 & -3.0525 & 0.3388 & -2.9721 & 0.3318 & +2.1 \\
 & -2.5037 & 0.2898 & -2.4822 & 0.2884 & +0.5 \\
 & -1.8695 & 0.2315 & -1.6316 & 0.2085 & +11.0 \\
 & -1.4228 & 0.1883 & -1.5927 & 0.2048 & -8.1 \\
 & -1.3333 & 0.1796 & -1.5927 & 0.2048 & -12.3 \\
\midrule
Type-BII
 & -4.2230 & 0.3913 & -4.0897 & 0.4255 & -8.0 \\
 & -4.0068 & 0.3739 & -4.0897 & 0.4255 & -12.1 \\
 & -3.4190 & 0.3700 & -3.2901 & 0.3591 & +3.0 \\
 & -2.4142 & 0.2823 & -2.4822 & 0.2884 & -2.1 \\
 & -2.3691 & 0.2782 & -2.4822 & 0.2884 & -3.6 \\
\bottomrule
\end{tabular}
}
\end{table}

\rev{The bare phase stiffness of the Type-BI and Type-BII arrays is compared with that of the Type-AI arrays over comparable Bloch bandwidths and is summarised in Table~3. Unlike the pronounced differences observed in the model-inferred effective phase stiffness, $K_{\mathrm{eff}}$, including the unphysical divergence obtained for the Type-BI arrays, the corresponding variations in the microscopic phase stiffness, $K_{0}$, are comparatively modest. For the Type-BI arrays, the percentage difference in $K_{0}$ increases slightly towards the strong-coupling regime, whereas no systematic trend is evident for the Type-BII arrays. These observations further indicate that the distinct critical-voltage scaling exhibited by the nanogap devices cannot be explained solely by changes in the microscopic electrodynamic properties of the arrays. Instead, they are consistent with substantial disorder-induced renormalisation of the collective depinning response, with the effect being most pronounced for the Type-BI arrays.}  

\subsection{Conclusion}

Taken together, these results demonstrate that the critical-voltage scaling in one-dimensional Josephson junction arrays is remarkably robust against a wide range of fabrication-induced structural variations. Variations in evaporation rate, substrate type (e.g., intrinsic \(\mathrm{Si}\), \(\mathrm{Si_3N_4}\) (300~nm)/\(\mathrm{Si}\)), film thickness, number of junctions, and island size do not measurably alter the normalized critical-voltage scaling. Likewise, \textit{in situ} post-oxidation modifies the room-temperature junction resistance and \rev{renormalises the collective transport response while preserving the underlying scaling behaviour.}

\rev{By contrast, arrays containing nanoscale gaps exhibit pronounced changes in both the critical-voltage scaling and the extracted scaling parameters. Within the theoretical framework of Bard \emph{et al.}, these observations are consistent with the nanoscale gaps introducing enhanced junction-to-junction variations in the local junction parameters and the associated fluctuations in the quantum phase-slip amplitude, thereby modifying the collective pinning landscape. The comparatively small changes in the microscopic phase stiffness, $K_{0}$, together with the much larger changes in the model-inferred effective stiffness, $K_{\mathrm{eff}}$, further indicate that the observed depinning behaviour is strongly renormalised by disorder beyond that expected from the bare circuit parameters alone. These results establish critical-voltage scaling as a practical experimental observable for differentiating how microscopic disorder mechanisms renormalise the collective insulating state in one-dimensional Josephson junction arrays.}

\subsection{Acknowledgment}
This work was supported by the Australian Research Council (ARC) under the Discovery Project DP No. DP190103370. All devices in this study were fabricated at the University of New South Wales (UNSW) node of the Australian National Fabrication Facility (ANFF).

\printbibliography

\end{document}